\documentstyle[twoside,fleqn]{article}

\newcommand{\Title}{
A coherent approach to Spacetime Foam }
\newcommand{\Authors}{R. Garattini\up{a} }
\newcommand{\Addresses}{
\up a Facolt\`a di Ingegneria, Universit\`a di Bergamo, Viale
Marconi 5, 24044 Dalmine (Bergamo), Italy.}
\newcommand{\ReprintRequests}{E-mail: Garattini@mi.infn.it}
\newcommand{\Abstract}{A coherent superposition of $N$
 Schwarzschild wormholes is proposed as a model for spacetime foam.
Following the subtraction procedure for manifolds with boundaries,
we calculate by variational methods the Casimir energy. A proposal
for an alternative foamy model formed by $N$ Schwarzschild-Anti-de
Sitter wormholes is here considered. Finally, a conjecture about
the foam evolution is proposed.}
\newcommand{\KeyWords}{
Spacetime Foam; Black Holes; Quantum Gravity.
}
\newcommand{\kopf}{
{\large\bf\Title} \vspace{3mm}\par \Authors \vspace{3mm}\par
\Addresses \vspace{3mm}\par \ReprintRequests
\vspace{3mm}\par
\Abstract
\vspace{3mm}\par
\KeyWords
\vspace{9mm}\par
}
\newcommand{\be}{\begin{equation}}
\newcommand{\ee}{\end{equation}}
\newcommand{\ba}{\begin{array}}
\newcommand{\ea}{\end{array}}

\newcommand{\up}[1]{$^{\rm #1}$}

\begin{document}

\sloppy
\thispagestyle{empty} \twocolumn[\kopf]

The term {\it Spacetime Foam} was used for the first time by J.A. Wheeler to
indicate that spacetime may be subjected to quantum fluctuations in topology
and metric at the Planck scale \cite{1}. In a series of papers we have
proposed a model for such a{\it \ foamy space} made by $N$ coherent
Schwarzschild wormholes \cite{2,3,4}. The relevance of this proposal is
based on a Casimir energy computation showing that quantum fluctuations of
the gravitational metric shift the minimum of the effective energy from flat
space ( the classical minimum of the energy) to a multi-wormhole
configuration. Indeed the final energy contribution to one loop is
\begin{equation}
\Delta _{N_{w}}E\left( M\right) \sim -N_{w}^{2}\frac{V}{64\pi ^{2}}\frac{%
\Lambda ^{4}}{e},  \label{a1}
\end{equation}
where $V$ is the volume of the system, $\Lambda $ is the U.V. cut-off and $%
N_{w}$ is the wormholes number \cite{2}. This expression shows
that a non-trivial vacuum of the multi-wormhole type is favoured
with respect to flat space. It is important to remark that it is
the $N$ - coherent superposition of wormholes that it is
privileged with respect to flat space and not the single wormhole,
because the single wormhole energy contribution has an imaginary
contribution in its spectrum: a clear sign of an instability.
Nevertheless the presence of an unstable mode is necessary to have
transition from one vacuum (the false one) to the other one (the
true vacuum) \cite{5} . Three consequences of this multiply
connected spacetime are:

\begin{enumerate}
\item  the event horizon area of a black hole is quantized and by means of
the Bekenstein-Hawking relation \cite{6,7}, also the entropy of a
black hole is quantized. In particular for a Schwarzschild black
hole
\begin{equation}
M=\frac{\sqrt{N}}{2l_{p}}\sqrt{\frac{\ln 2}{\pi }},
\end{equation}
namely the black hole mass is quantized. Here $l_{p}$ is the Planck length
in natural units.

\item  A cosmological constant is induced by vacuum fluctuations as shown by
$\left( \ref{a1}\right) $ whose value is
\begin{equation}
\Lambda _{c}=\frac{\Lambda ^{4}l_{p}^{2}}{N_{w}8e\pi }.  \label{a2}
\end{equation}
When the area-entropy relation is applied to the de Sitter geometry, we
obtain
\begin{equation}
\frac{3\pi }{\ln 2l_{p}^{2}N_{w}}=\Lambda _{c}.  \label{a3}
\end{equation}

\item  Combining $\left( \ref{a2}\right) $ and $\left( \ref{a3}\right) $,
one gets
\begin{equation}
\Lambda _{c}=\frac{\Lambda ^{4}l_{p}^{2}}{N_{w}8e\pi }=\frac{3\pi }{\ln
2l_{p}^{2}N_{w}}.
\end{equation}
This means that we have found a constraint on the U.V. cut-off
\begin{equation}
\Lambda ^{4}=\frac{24e\pi ^{2}}{\ln 2l_{p}^{4}}.
\end{equation}
\end{enumerate}

It is interesting to note that $\left( \ref{a1}\right) $ can be obtained at
least for $N_{w}=1$ even for Schwarzschild-Anti-de Sitter wormholes (S-AdS)
\cite{8}, whose line element is $ds^{2}=$%
\begin{equation}
-f\left( r\right) dt^{2}+f\left( r\right) ^{-1}dr^{2}+r^{2}d\Omega ^{2},
\end{equation}
where
\begin{equation}
f\left( r\right) =\left( 1-\frac{2MG}{r}+\frac{r^{2}}{b^{2}}\right) .
\label{a4}
\end{equation}
$\Lambda _{AdS}$ is the negative cosmological constant and $b=\sqrt{-\frac{3%
}{\Lambda _{AdS}}}$. To consider a large $N_{w}$ approach to spacetime foam
even with S-AdS wormholes, one has to consider the following rescaling
\begin{equation}
\left\{
\begin{array}{c}
R_{\pm }\rightarrow R_{\pm }/N_{w} \\
l_{p}^{2}\rightarrow N_{w}l_{p}^{2} \\
\Lambda _{AdS}\rightarrow \Lambda _{AdS}/N_{w}^{2}
\end{array}
\right. ,  \label{a5}
\end{equation}
where $R_{\pm }$ are the boundaries related to the single wormhole. This
rescaling is a consequence of the boundary reduction related to the coherent
superposition of wormholes wave functionals leading to the stabilization of
the system This means that a selection rule has to emerge to compare the
quantity
\[
\Gamma _{{\rm N-S-AdS\ holes}}
\]
\begin{equation}
=\frac{P_{{\rm N-S-AdS\ holes}}}{P_{{\rm AdS}}}\simeq \frac{P_{{\rm foam}}}{%
P_{{\rm AdS}}}  \label{a6}
\end{equation}
with
\[
\Gamma _{{\rm N-S}\ {\rm holes}}
\]
\begin{equation}
=\frac{P_{{\rm N-S}\ {\rm holes}}}{P_{{\rm flat}}}\simeq \frac{P_{{\rm foam}}%
}{P_{{\rm flat}}}.  \label{a7}
\end{equation}
In both cases, we find a non-vanishing probability that a non-trivial vacuum
has to be considered. Moreover the rescaling in $\left( \ref{a5}\right) $
leaves the potential $\left( \ref{a4}\right) $ invariant and when $N_{w}$ is
very large $\Lambda _{AdS}\rightarrow 0$. This seems to suggest that not
only $\left( \ref{a6}\right) $ and $\left( \ref{a7}\right) $ have to be
compared, but it is likely that a hierarchical mechanism of the type
\begin{equation}
\begin{array}{c}
\frac{P_{{\rm N-S-AdS\ holes}}}{P_{{\rm AdS}}}\simeq \frac{P_{{\rm foam}}}{%
P_{{\rm AdS}}} \\
\downarrow _{\Lambda _{AdS}\rightarrow 0} \\
Flat\quad Space \\
\downarrow \\
\frac{P_{{\rm N-S}\ {\rm holes}}}{P_{{\rm flat}}}\simeq \frac{P_{{\rm foam}}%
}{P_{{\rm flat}}}
\end{array}
,  \label{a8}
\end{equation}
when $N_{w}\rightarrow \infty $. Although this picture has to be examined in
detail and $\left( \ref{a8}\right) $ is only at the conjecture level, an
important question comes into play: why a spacetime formed by wormholes has
to be preferred with respect to flat spacetime, when the last one is the one
we observe. The answer that at this stage can only be conjectured is that if
we consider the following expectation value on the foam state
\begin{equation}
\frac{\left\langle \Psi _{F}\left| \hat{g}_{ij}\right| \Psi
_{F}\right\rangle }{\left\langle \Psi _{F}|\Psi _{F}\right\rangle },
\end{equation}
when the number of wormholes is large enough, i.e. the scale is sufficiently
large, we should have to obtain
\begin{equation}
\frac{\left\langle \Psi _{F}\left| \hat{g}_{ij}\right| \Psi
_{F}\right\rangle }{\left\langle \Psi _{F}|\Psi _{F}\right\rangle }%
\rightarrow \eta _{ij},
\end{equation}
where $\eta _{ij}$ is the flat space metric. This is a test that this foamy
model has to pass if phenomenological aspects have to be considered.

\bigskip {\bf Acknowledgments}

\smallskip \noindent I would like to thank Prof. R. Bonifacio who has given
to me the opportunity of participating to the Conference. 


\bigskip

\begin{thebibliography}{9}
\bibitem{1}  J.A. Wheeler, Ann. Phys. {\bf 2 }(1957) 604; J.A. Wheeler, {\it %
Geometrodynamics}. Academic Press, New York, 1962.

\bibitem{2}  R. Garattini, {\it A Spacetime Foam approach to the
cosmological constant and entropy}. To appear in Int.J.Mod.Phys. {\bf D};
gr-qc/0003090.

\bibitem{3}  R. Garattini, Phys. Lett. {\bf B 446} (1999) 135,
hep-th/9811187.

\bibitem{4}  R. Garattini, Phys. Lett. {\bf B 459} (1999) 461,
hep-th/9906074.

\bibitem{5}  S. Coleman, Nucl. Phys. {\bf B} {\bf 298} (1988) 178.

\bibitem{6}  J. Bekenstein, Phys. Rev. {\bf D 7} (1973) 2333.

\bibitem{7}  S. Hawking, Phys. Lett. {\bf B 134} (1984) 403.

\bibitem{8}  R. Garattini, Class.Quant. Grav.{\bf \ 17} (2000) 3335,
gr-qc/0006076.
\end{thebibliography}
\end{document}